\documentclass[aps,pra,twocolumn,groupedaddress]{revtex4}
\usepackage{epsfig}
\usepackage{graphicx}
\usepackage{amsfonts}
\usepackage{amsmath,amssymb}
\usepackage{latexsym}
\usepackage{bm}

\newcommand{\beq}{\begin{eqnarray}}
\newcommand{\eeq}{\end{eqnarray}}
\newcommand{\nn}{\nonumber}
\newcommand{\hb}{\hbar}
\newcommand{\8}{\infty}
\newcommand{\f}{\frac}
\newcommand{\rfs}[1]{Eq.~(\ref{#1})}
\newcommand{\rf}[1]{\ref{#1}}
\newcommand{\h}[1]{\hat{#1}}
\newcommand{\dr}[1]{|#1\rangle}
\newcommand{\dl}[1]{\langle#1}
\newcommand{\tb}[1]{\textbf{#1}}
\newcommand{\tr}[1]{\textrm{#1}}

\def\nn{\nonumber}

\def\e{\epsilon}
\def\d{\delta}
\def\L{\Lambda}
\def\o{\omega}
\def\B{\beta}
\def\G{\Gamma}
\begin{document}
\title{Correlation effects in ultracold two-dimensional Bose gases}
\author{Lih-King Lim, C. Morais Smith, and H. T. C. Stoof}
\affiliation{Institute for Theoretical Physics, Utrecht
University, Leuvenlaan 4,3584 CE Utrecht, The Netherlands}
\date{\today}
\begin{abstract}
We study various properties of an ultracold two-dimensional (2D)
Bose gas that are beyond a mean-field description. We first derive
the effective interaction for such a system as realized in current
experiments, which requires the use of an energy dependent
$T$-matrix. Using this result, we then solve the mean-field
equation of state of the modified Popov theory, and compare it
with the usual Hartree-Fock theory. We show that even though the
former theory does not suffer from infrared divergences in both
the normal and superfluid phases, there is an unphysical density
discontinuity close to the Berezinskii-Kosterlitz-Thouless
transition. We then improve upon the mean-field description by
using a renormalization group approach and show how the density
discontinuity is resolved. The flow equations in two dimensions, in particular,
of the symmetry-broken phase, already contain some unique features
pertinent to the 2D $XY$ model, even though vortices have not been
included explicitly. We also compute various many-body
correlators, and show that correlation effects beyond the
Hartree-Fock theory are important already in the normal phase as
criticality is approached. We finally extend our results to the
inhomogeneous case of a trapped Bose gas using the local-density
approximation and show that close to criticality, the
renormalization group approach is required for the accurate
determination of the density profile.
\end{abstract}

\maketitle
\section{Introduction}
Low-dimensional systems play a unique role in the study of
many-body effects. For example, the enhanced importance of thermal
fluctuations prevents a two-dimensional (2D) system with a
continuous symmetry to undergo spontaneous symmetry breaking at
any nonzero temperature, thus preventing the presence of true
long-range order. This property is elucidated in the
Mermin-Wagner-Hohenberg theorem \cite{Mermin:66,Hohenberg:67}.
Nonetheless, the system still exhibits interesting properties. In
particular, the 2D $XY$ model undergoes a special type of phase
transition into a state which is characterized by only algebraic
long-range order instead. The underlying mechanism that drives the
phase transition, known as the Berezinskii-Kosterlitz-Thouless
(BKT) transition \cite{Bere:72,Kost:73}, is the unbinding of
vortex-antivortex pairs. Due to its topological nature, such a
phase transition is difficult to incorporate into the standard
Ginzburg-Landau theory with a local order parameter. For the
ultracold 2D Bose gas, the absence of Bose-Einstein condensation
(BEC) requires the concept of a quasi-condensate to understand the
existence of algebraic long-range order and superfluidity
\cite{Popov:83,Shlyapnikov:00,Kagan:00,Stoof:02,Hutchinson:04}.
Experiments in the field of ultracold atomic gases have recently
reached this interesting 2D regime to allow for the direct
observation of this phenomenon in a highly controllable
environment \cite{Dalibard:05,Dalibard:06,Cornell:07,Dalibard:07,Phillips:08}.
The observation of dislocations in the interference pattern of two
condensates \cite{Walls:98,Devreese:98,Dalibard:05,Cornell:07} and
the studies of the coherence properties \cite{Demler:06,
Dalibard:06}, for example, have all agreed with the universal
predictions of BKT theory. More experiments and numerical
simulations have come to address also various nonuniversal
properties specific to the atomic gases under consideration
\cite{Dalibard:07,Dalibard:07b,Krauth:07,Krauth:08,Blakie:08}.

To describe the low-dimensional Bose gas at nonzero temperatures,
the usual approach of the Bogoliubov theory is plagued with
infrared divergences. However, these divergences can be shown to
occur due to a spurious contribution from the condensate phase
fluctuations in the Bogoliubov approach and by removing these
contributions we can arrive at a modified Popov theory, which is
valid for any dimension and at all temperatures \cite{Stoof:02}.
In this paper, we first study this modified Popov theory for the
ultracold 2D Bose gas and show that it contains a density
discontinuity above the BKT transition, close to the point where
the quasicondensate density becomes nonzero. To improve upon the
mean-field description, we next employ a renormalization group
(RG) approach to take into account the quantum and thermal
fluctuations more accurately, in particular in the normal phase,
when the modified Popov theory reduces to Hartree-Fock theory. The
RG theory developed for the ultracold three-dimensional (3D) Bose
gas was shown to be quantitatively successful in addressing
effects beyond mean-field \cite{Stoof:96}. Here, we derive the
analogous 2D RG flow equations and find the surprising result that
they show features which are very different from the 3D RG theory.
We first show how various characteristics of a quasicondensate
are manifested in this framework. We then interpret these unique
features as precursors of the BKT physics, even though we have not
taken topological defects into account explicitly. This is because
the RG theory is derived from the full atomic quantum field
$\psi(\tb{x},\tau)$ and not from its phase alone \cite{Wetterich:08}. With the RG
approach, the density correction to the Hartree-Fock theory indeed
resolves the unphysical density discontinuity in the mean-field
description. Furthermore, it agrees with the equation of state of
the modified Popov theory already above the critical temperature,
which shows that the latter has correctly included correlation
effects beyond the Hartree-Fock description even in the normal
phase.

The paper is organized as follows. In Sec. \rf{sec:T}, we discuss
the exact form of the $T$-matrix for ultracold 2D gases as
realized under current experimental conditions, since it
determines the effective interaction of the 2D Bose gas. Next, in
Sec. \rf{sec:mf}, we present the mean-field results from the
modified Popov theory, and point out that the Hartree-Fock theory
becomes unstable already above the BKT transition, even though a
solution to the Hartree-Fock equation of state exists at all
temperatures. The discontinuity in the density close to the BKT
transition that results from the modified Popov theory is deemed
unsatisfactory, and in Sec. \rf{sec:RG}, we present the required
RG theory to improve upon this result. We discuss various
interesting features of the flow equations in two dimensions and compute
various nonuniversal quantities of interest within this approach.
While the RG approach resolves the density discontinuity of the
mean-field equation of state, it remains incapable of capturing
all the critical properties known from the BKT physics. We also
compute various many-body correlators, and show how correlation
effects beyond a mean-field picture show up in the normal phase
close to criticality. In Sec. \rf{sec:dis}, we extend our results to
the case of a trapped Bose gas and compare them with the ideal gas.
We end with some concluding remarks in Sec. \ref{sec:con}.\\

\section{Effective interaction in the 2D regime}
\label{sec:T} Even though we shall be interested in ultracold 2D
Bose gases, in experiments with atomic alkali-metal gases the
realization of such a system is achieved by restricting the motion
of a trapped 3D gas onto a plane. For the quantum degenerate gas
to be in the 2D regime, one needs to ensure that the motion along
the tightly confining axial direction is frozen out. This
condition is met if $k_B T, \mu\ll \hb \o_z$, where  $k_B T$ is
the thermal energy, $\mu$ is the chemical potential, and $\o_z$ is
the axial trapping frequency.

In the ultracold limit of the gases under consideration, the
effective interaction is in the first instance determined by the
three-dimensional two-body $T$-matrix. We therefore begin with
considering the full two-body Hamiltonian $\h{H}_0+\h{V}$ in the
center-of-mass coordinate frame, where the free Hamiltonian
$\h{H}_0$ is given by \beq \h{H}_0=
-\f{\hb^2}{m}\vartriangle_{\tb{r}}-\f{\hb^2}{m}\vartriangle_{z}+\f{1}{4}m\o_z^2z^2,
\eeq $m$ is the atomic mass, and $\h{V}$ is the interaction
potential modeled by a short-ranged delta function of strength
$V_0$. The center-of-mass coordinate frame is spanned by the
$\tb{r}$-plane, which is taken to be homogenous, and the $z$-axis
in the tightly-confining direction. We denote the eigenstates of
$\hat{H}_0$ by  $\dr{\tb{k},n}$, which are given by the product of
2D plane waves $\dl{\tb{r}}\dr{\tb{k}}=e^{i \tb{k} \cdot \tb{r}}$
and one-dimensional oscillator functions $\dl{z}\dr{n}=\phi_n(z)$.
The latter functions are defined in the standard form,
$\phi_n(z)=(2\pi)^{-1/4}(2^n n! \, l)^{-1/2} H_n(z/ \sqrt{2}l)
\exp(-z^2/4l^2)$, where $l=\sqrt{\hb/m\o_z}$ is the harmonic
length in the axial direction and $H_n(z)$ is the Hermite
polynomial.

The Lippmann-Schwinger equation for the two-body $T$-matrix is
given by \beq \label{LSeqn}
\h{T}(E^+)=\h{V}+\h{V}\f{1}{E^+-\h{H}_0}\,\h{T}(E^+), \eeq where
$E^+=E+i0$ and the effective interaction in the 2D regime is given
by the matrix element with respect to the axial ground state,
i.e., \beq T_{00}(E)\equiv\dl{\tb{k},n=0} |\h{T}(E)
\dr{\tb{k}',n=0}. \eeq By inserting the completeness relation
$\sum_n \int d \tb{k}  \dr{\tb{k},n}\dl{\tb{k},n}|/(2 \pi)^2=1$,
the operator equation in \rfs{LSeqn} can be written as \beq
\label{tmat} \frac{1}{T(E)}&=&\frac{1}{V_0}-\sum_{n=0}^{\infty}
\int \frac{d^2 \tb{k}}{(2 \pi)^2}
\frac{|\phi_n(0)|^2}{E-E_0(\tb{k},n)}, \eeq where $T(E)$ is the
matrix element with respect to 3D plane waves, which is related to
the desired quantity $T_{00}(E)$ by means of
$T_{00}(E)=|\phi_0(0)|^2T(E)$. Moreover, $E_0(\tb{k},n)$ are the
eigenvalues of the free Hamiltonian \beq E_0(\tb{k},n)=2
\epsilon_{\tb{k}}+ \left(n+\f{1}{2}\right)\hb \omega_z, \eeq with
$\epsilon_{\tb{k}}=\hb^2 \tb{k}^2/2m$.

As it stands, \rfs{tmat} suffers from an ultraviolet divergence due to the delta function interaction potential, which neglects any momentum dependence at high momentum. To cure this divergence, we observe that while the harmonic oscillator functions with odd $n$ vanish at the origin, the functions with even $n$ have the asymptotic behavior
\beq
|\phi_{2n}(0)|^2=\f{1}{\pi l\sqrt{2}}\f{\Gamma(n+\f{1}{2})}{\Gamma(n+1)}\sim\f{1}{\pi l\sqrt{2n}}+ {\cal O}\, \left(\f{1}{n^{3/2}}\right),
\eeq
for large quantum numbers $n$, where $\Gamma(n)$ denotes the Gamma function. It then follows that, for large $n$, the second term on the right-hand side of \rfs{tmat} takes the form
\beq
\int \f{d^2\tb{k}dk_z}{(2 \pi)^3}\f{1}{E -2\epsilon_{\tb{k}}-2\epsilon_{k_z}},
\eeq
with $k_z\equiv\sqrt{2 n}/l$. This form diverges in the ultraviolet in exactly the same manner as the two-body $T$-matrix in the homogenous three-dimensional space, i.e., as
\beq
\frac{1}{T^{2B}(E)}&=&\frac{1}{V_0}-\int \frac{d^3 \tb{q}}{(2 \pi)^3} \frac{1}{E-2 \epsilon_{\tb{q}}},
\eeq
with $\tb{q}=(\tb{k},k_z)$. On physical grounds, this behavior is expected since the short-distance behavior of the gas is not altered by the harmonic confinement and a standard renormalization procedure can therefore be implemented. The latter is achieved by eliminating the bare coupling strength $V_0$ in favor of the relevant physical parameter $T^{2B}(E\simeq0)= 4 \pi \hb^2 a/m$ for the low-energy scattering, where $a$ is the 3D scattering length. After the elimination, a well-defined expression for the $T$-matrix,
\beq\label{rTmat}
\frac{1}{T(E)}&=&\f{m}{4 \pi \hb^2 a}-\sum_{n=0}^{\infty} \int \frac{d^2 \tb{k}}{(2 \pi)^2}  \frac{|\phi_n(0)|^2}{E-E_0(\tb{k},n)}\nn\\&&-\int \frac{d^3 \tb{q}}{(2 \pi)^3} \frac{1}{2 \epsilon_{\tb{q}}},
\eeq
that no longer contains any divergence is obtained. It is shown in the Appendix that the effective interaction can be further worked out to yield
\beq\label{rTmat2}
T_{00}(E)=\f{2 \sqrt{2 \pi} \hb^2}{m}\,\f{1}{l/a+F(E) },
\eeq
with
\beq
F(E)=F_0\!+\!\f{1}{\sqrt{2 \pi}}\!\sum_{n=0}^{\8}\! \f{(2n-1)!!}{(2n)!!} \ln \!\left[ \f{4n+1}{4n -2E/\hb \omega_z}\right]\!\!,
\eeq
where $F_0\simeq -0.3508$, and for convenience energies are measured with respect to the zero-point energy $\hb \o_z/2$ of the axial ground state. This result can be compared with the previous work of Petrov \textit{et al.} \cite{Shlyapnikov:01}, where a different approach has been adopted. Thus, we see that the interaction takes a quasi-2D form, even though the system is kinematically two dimensional. The $T$-matrix $T_{00}(E)$ interpolates between the form of the interaction for the 3D and 2D cases, where the former is a constant and the latter has a logarithmic energy dependence. For the trapped atomic gases of interest, the condition $l/a\gg1$ is usually satisfied. Hence the effective interaction is often well approximated by $T_{00}(E) \simeq 2 \sqrt{2 \pi}a\hb^2/ml\equiv g$. Nonetheless, the exact form of the $T$-matrix with an energy dependence will turn out to be important in the present work. In Fig. \rf{Fig.1}, we show $|T_{00}(E)|$ as a function of energy. Besides that the overall scale is set by $g$, there are zeros at energies $2n\hb\o_z$ corresponding to the harmonic oscillator states with even $n$, due to the divergence of the logarithmic function. Note again that we have subtracted the zero-point energy.

\begin{figure}[t]
\includegraphics[scale=.68, angle=0, origin=c]{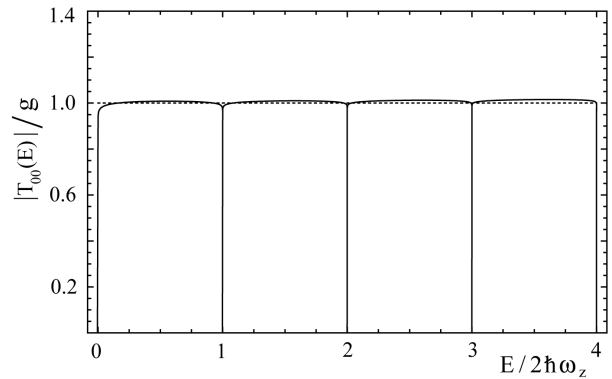}
\caption{\label{Fig.1} The exact $T$-matrix element $|T_{00}(E)|$ as a function of energy.}
\end{figure}

\section{Mean-Field Theory}\label{sec:mf}
In this section, we recapitulate the main result obtained from the modified Popov theory for the ultracold 2D Bose gas \cite{Stoof:02}. The equation of state in this mean-field theory is given by
\beq\label{sf}
n\negthickspace\!&=&\!\negthickspace n_0\!+\!\f{1}{V}\!\sum_{\tb{k}}\! \left\{ \f{\epsilon_{\tb{k}}}{2 \hb \omega_{\tb{k}}} \left[2 N(\hb \o_{\tb{k}})+1\right]\!-\!\f{1}{2}\!+\!\f{n_0 T_{00}(-2\mu)}{2 \epsilon_{\tb{k}}+2 \mu} \right\}\nn\\
\negthickspace\!&\equiv&\!\negthickspace n_0\!+\! n'
\eeq
and $\mu=(2n-n_0)\,T_{00}(-2 \mu)$,
where $n$ is the total density, $n_0$ is the (quasi)condensate density, $n'$ is the density fluctuations around $n_0$, $\hb \o_{\tb{k}}=[\e^2_{\tb{k}}+2n_0 T_{00}(-2\mu)\e_{\tb{k}}]^{1/2}$ is the Bogoliubov quasiparticle dispersion, $N(x)=1/(e^{\B x}-1)$ is the Bose-Einstein distribution function, and $\B=1/k_B T$ is the inverse thermal energy. Notice that these equations do not suffer from infrared and ultraviolet divergences, and thus, they are valid for any dimension and at all temperatures. This comes about because the condensate phase fluctuations have been treated exactly to arrive at this result. A condensate with true long-range order is absent in two dimensions at any nonzero temperature. In fact, as shown in Ref. \cite{Stoof:02}, $n_0$ should be identified with the quasicondensate density at any nonzero temperature. Due to its mean-field nature, however, this equation of state is incapable to capture the BKT transition. Indeed, although the criterion for the BKT transition is $n_s \Lambda^2_{\textrm{th}}=4$, where $\Lambda_{\textrm{th}}=\sqrt{2\pi\hb^2/m k_B T}$ is the thermal de Broglie wavelength of the atoms in the gas, according to \rfs{sf}, a nontrivial solution exists even if the superfluid density $n_s$ obeys $n_s \Lambda^2_{\textrm{th}}<4$. To circumvent this shortcoming, it was shown that by complementing the above equation of state with a renormalization group analysis on the Sine-Gordon model that takes into account vortices explicitly, the canonical criterion for the BKT transition is met when $n_0 \Lambda^2_{\textrm{th}}= 6.65$ \cite{Stoof:02}. Thus, above this critical temperature, the fugacity of the vortices renormalizes at long wavelengths to a nonzero value, which leads to the destruction of superfluidity but not immediately of the quasi-condensate.

The normal state without a quasicondensate is instead described
by the Hartree-Fock equation of state given by \beq\label{HF}
n=\f{1}{V}\sum_{\tb{k}} N(\epsilon_{\tb{k}}+\hb \Sigma -\mu), \eeq
where the Hartree-Fock self-energy $\hb \Sigma$ satisfies \beq \hb
\Sigma=2nT_{00}(-\hb \Sigma). \eeq It is also noted that when the
$T$-matrix is energy independent, i.e., $T_{00}(E)\simeq g$, the
normal equation of state can actually be satisfied at all
temperatures, since \rfs{HF} can then be integrated to give
\beq\label{hf1} n
\Lambda^2_{\textrm{th}}=-\ln(1-e^{-\beta(2ng-\mu)}). \eeq For a
fixed total density $n$, \rfs{hf1} has always a solution for the
chemical potential $\mu$, at all temperatures. Thus to ensure the
stability of the normal phase, we have to examine with the same
chemical potential $\mu$ whether the solution of the modified
Popov theory exists.

To illustrate this discussion, we numerically solve both equations of state, as
shown in Fig. \rf{Fig.2}. We observe that there is an unphysical
discontinuity in the density curve due to a mismatch in the total
density between the two equations of state close to the BKT
transition. This is an artifact of the mean-field theory which
will be resolved in the next section.

\begin{figure}
\includegraphics[scale=.7, angle=0, origin=c]{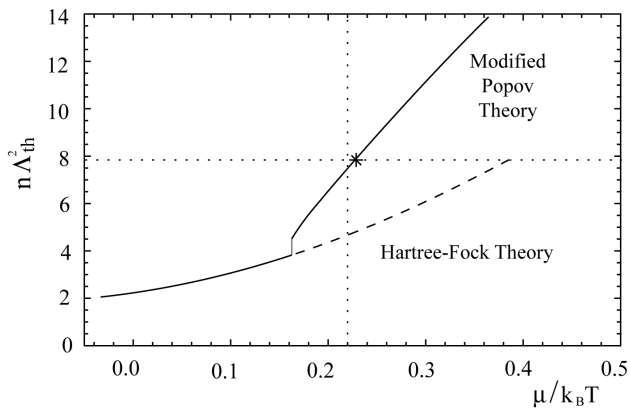}
\caption{\label{Fig.2} The mean-field equation of state from the
modified Popov theory. The dashed curve is the Hartree-Fock theory
that is extended also into the unstable regime until the
Monte Carlo critical condition $n_c \L_{\textrm{th}}^2=7.81$ is met
(dotted line). We have taken $m g/\hb^2=0.15$, which roughly
corresponds to the experiment of Ref. \cite{Dalibard:05}. The
canonical BKT criterion derived from the modified Popov theory is
indicated with an asterisk.}
\end{figure}

To compare, we include in Fig. \rf{Fig.2} the result from the
usual Hartree-Fock mean-field approach. In this approach, the
Hartree-Fock theory is employed up to a chemical potential that
corresponds to the critical density $n_c$ obtained from the high
precision Monte Carlo simulations \cite{Proko:01}. Numerical
simulations yielded a critical density for the onset of the BKT
transition of $n_c \Lambda^2_{\textrm{th}}\simeq \ln(380 \,\hb^2/m
g)=7.81$, and a critical chemical potential $\mu_c/k_B
T\simeq(mg/\pi\hb^2) \ln(13.2\,\hb^2/mg)=0.22$, for
$mg/\hb^2\simeq0.1541$. Both conditions are shown by the dotted
lines in Fig. \rf{Fig.2}. It is noted that the critical chemical potential
obtained within this mean-field approach, $\mu_c^{HF}/k_B
T\simeq0.38$, differs from the Monte-Carlo result, thus indicating
an inconsistency in the Hartree-Fock approach. This can be understood
within the modified Popov theory as an instability of the system
toward a more correlated phase. On the other hand, despite the
density discontinuity, the modified Popov theory yields a critical
density $n_c^{MP} \Lambda^2_{\textrm{th}}\simeq 7.94$ and a
critical chemical potential $\mu_c^{MP}/k_B T\simeq0.23$, which
are in excellent agreement with the Monte Carlo results. This
shows that the main problem of the modified Popov theory lies in
the inaccurate treatment of correlations in the region where it
reduces to Hartree-Fock theory.

\begin{figure}
\includegraphics[scale=.53, angle=0, origin=c]{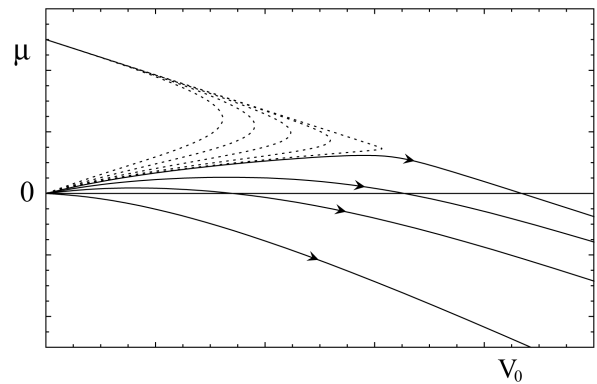}
\caption{\label{Fig.3} The RG flow diagram of $V_0$ and $\mu$ in the symmetric phase. The arrows indicate the flow direction towards the long-wavelength regime. The dotted lines are ill-defined trajectories as explained in the text.}
\end{figure}

\section{Renormalization Group Theory}\label{sec:RG}
To improve upon the modified Popov theory, we employ here an RG approach \cite{Wilson:74}. In this approach, we systematically integrate out the high-momentum shell $\L e^{-t}<k<\L$, and absorb its contribution into the parameters of the theory, which then become dependent on the flow parameter $t$. Here, $\L$ is the ultraviolet cutoff of the theory that is specified below. As discussed in Ref. \cite{Stoof:96}, due to the ultracold limit of the Bose gas under consideration, the parameters which are important in determining the various properties of interest are the chemical potential $\mu=\mu(t)$, and the two-body interaction strength $V_0=V_0(t)$.

\subsection{The Flow Equations}\label{subsec1:RG}
We shall present here the flow equations for these parameters in two dimensions, thus extending the results obtained in Refs. \cite{Stoof:96,note1}. In the symmetric phase, for $\mu<0$, the flow equations are given by
\beq\label{RGnormal}
\f{d\mu}{dt}&=&2\mu-\f{\L^2}{\pi}V_0N(\e_{\L}-\mu),\nn\\
\f{d V_0}{dt}&=&-\f{\L^2}{2 \pi} V_0^2  \left[\f{1+2 N(\e_{\L}-\mu)}{2(\e_{\L}-\mu)}+4\B N(\e_{\L}-\mu)\right.\nn\\&&\times[N(\e_{\L}-\mu)+1]\biggr],
\eeq
while in the symmetry-broken phase, for $\mu>0$, they are
\beq\label{RGsf}
\f{d\mu}{dt}&=&2\mu-\f{\Lambda^2}{2 \pi}V_0\left[\f{2 \epsilon^3_{\Lambda}+6\mu\epsilon^2_{\Lambda}+\mu^3}{2 \hb^3 \omega^3_{\Lambda}}[2 N(\hb \omega_{\Lambda})+1]-1\right.\nn\\&&\left.+\f{\mu (2\epsilon_{\Lambda}+\mu )^2}{\hb^2 \omega^2_{\Lambda}}\beta N(\hb \omega_{\Lambda})[N(\hb \omega_{\Lambda})+1]\right],\nn\\
\f{dV_0}{dt}&=&-\f{\Lambda^2}{2 \pi}V_0^2 \left[ \f{(\epsilon_\Lambda-\mu)^2}{2 \hb^3 \omega^3_{\Lambda}}[2 N(\hb \omega_{\Lambda})+1]\right.\nn\\&&\left.+\f{(2\epsilon_{\Lambda}+\mu )^2}{\hb^2 \omega^2_{\Lambda}}\beta N(\hb \omega_{\Lambda})[N(\hb \omega_{\Lambda})+1]\right].
\eeq
Here, the chemical potential has been trivially rescaled by $\mu\rightarrow\mu e^{2 t}$, hence the $2\mu$ term in the flow equations. The inverse temperature has also been trivially rescaled as $\B \rightarrow\B e^{-2 t}$, but the interaction strength does not scale trivially in two dimensions. To examine the critical properties, however, a different trivial scaling needs to be used. This comes about because close to the critical regime, where the correlation length and correlation time diverge, the time-derivative term in the quantum action can be neglected with respect to the kinetic term. This is equivalent to taking the large-$t$ or high-temperature limit of the flow equations by setting $N(x)\rightarrow1/\B x$. As a result, while the trivial scaling of the chemical potential remains the same, the coupling strength acquires a trivial scaling with exponent 2, i.e., $V_0\rightarrow V_0 e^{2 t}$. As expected, the trivial scalings in the critical regime then agree with the trivial scalings of the classical $2D$ $XY$ model.

\begin{figure}
\includegraphics[scale=.58, angle=0, origin=c]{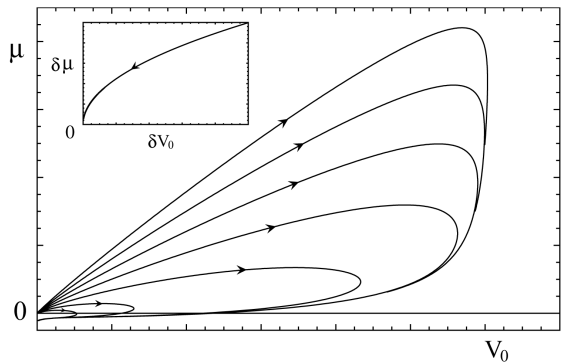}
\caption{\label{Fig.4} The RG flow diagram in the symmetry-broken phase. There is a limiting curve which ends at the attractor $(0,-\e_{\L}/2)$ where all trajectories approaches asymptotically. The inset shows the approach to the attractor.}
\end{figure}

To solve the flow equations, the correct boundary conditions have to be provided. The initial value for the chemical potential $\mu(t=0)$ is nothing but the bare chemical potential in the original theory. Its initial value can either be positive or negative, corresponding to an initial phase which is symmetry broken or unbroken, respectively. Furthermore, the flow equations allow the chemical potential to change sign, since the two sets of equations are smoothly connected at $\mu=0$. As we shall see, it is in fact a general feature of the solutions to the 2D flow equations in the symmetry-broken phase that the chemical potential always flows to a negative value for large $t$, which is not the case in three dimensions.

To determine the boundary condition for the interaction strength, we recognize that in vacuum, i.e., by setting $N(\e_{\L}-\mu)=0$, the flow equation for the interaction strength
\beq\label{eq18}
\f{d V_0}{dt}=-\f{\L^2}{2 \pi}\f{V_0^2}{2(\e_{\L}-\mu)}
\eeq
is nothing but the differential form of the Lippmann-Schwinger equation for the $T$-matrix at energy $2\mu$. Since the $T$-matrix solution to the Lippmann-Schwinger equation, as obtained in Sec. \ref{sec:T}, entails the summation of all ladder diagrams for the scattering process in vacuum, we have to ensure that the flow equation in \rfs{eq18} reproduces the correct long-wavelength result for large $t$. In particular, the initial value $V_0(t=0)$ is chosen such that for large $t$ the correct form of the $T$-matrix is recovered in the vacuum, i.e., $V_0(t\rightarrow \8)= T_{00}(2\mu)$. This can be satisfied by the following initial condition:
\beq
V_0(t=0)=\f{2\sqrt{2 \pi}\hb^2}{m}\f{1}{l/a+(1/\sqrt{2 \pi})\ln\left( m\o_z/2\hb\L^2\right)},
\eeq
for $\hb^2 \L^2/m  \gg 2\mu$. It is important to note that for $\L\gg \L_{\tr{th}}$, this procedure allows us to eliminate the ultraviolet cutoff dependence of the theory since the interaction strength already attained the two-body $T$-matrix value before entering the thermal regime \cite{Stoof:96}. With these boundary conditions and the ultraviolet cutoff chosen to be $\L\sim {\cal O}\, (100 \L_{\tr{th}})$, we now numerically integrate the flow equations to obtain the specific solution $\{V_0(t),\mu(t)\}$ for different chemical potentials.\\

\begin{figure}
\includegraphics[scale=1.2, angle=0, origin=c]{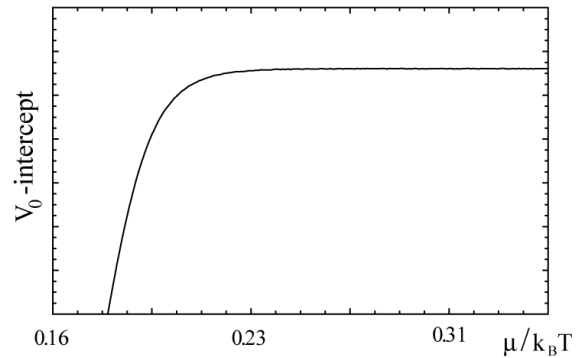}
\caption{\label{Fig.5a} The intercept with the $V_0$ axis as a function of the chemical potential $\mu$.}
\end{figure}

\subsection{Analysis of the Flow Equations}\label{subsec2}
We first study the two sets of flow equations in \rfs{RGnormal} and \rfs{RGsf} separately, and allow for the chemical potential in both equations to take positive and negative values. In the symmetric phase, as seen in Fig. \rf{Fig.3}, the trajectories seem to be characteristics of the order-disorder phase transition of the classical 3D $XY$ model. The fixed point of the flow equations can be found by considering their large-$t$ limit
\beq
\f{d\mu}{dt}&=&2\mu-\f{\L^2 V_0}{\pi}\f{k_B T}{\e_{\L}-\mu}=0,\nn\\
\f{d V_0}{dt}&=&2V_0-\f{\L^2 V_0^2}{2\pi}\f{5 k_B T}{(\e_{\L}-\mu)^2}=0,
\eeq
which yields $(V_0^*,\mu^*)=(20\pi  \e_{\L}^2/49 \L^2 k_B T,2\e_{\L}/7)$. However, it is important to note that for chemical potentials above a critical value, the resulting trajectories are actually ill-defined because $\mu(t)$ eventually grows to a point where the Bose-Einstein distribution diverges, i.e. $N(\e_{\L}-\mu)=\8$. Thus, there is not really an order-disorder phase transition with increasing chemical potential. More generally, the first quadrant with positive chemical potential should be regarded as the unphysical region of the symmetric phase because the true minimum of the action is shifted away from the origin. Similarly, in three dimensions the unstable fixed point lying in this region does not give the most accurate critical properties of the system \cite{Stoof:96}.

In the symmetry-broken phase, the flow equations present a few rather surprising features, as shown in Fig. \rf{Fig.4}. First, all trajectories always flow into the fourth quadrant with a negative chemical potential. This feature can be interpreted as the \textit{prima facie} property of a quasicondensate in two dimensions, where starting with an initial ``condensate'' $(\mu>0)$ at the shortest distance, as the high momentum shells are being integrated out, the symmetry always gets restored ($\mu<0$) in the infrared regime. In other words, the high-momentum fluctuations at long wavelength always destroy the coherence present at short distances.

\begin{figure}
\includegraphics[scale=.57, angle=0, origin=c]{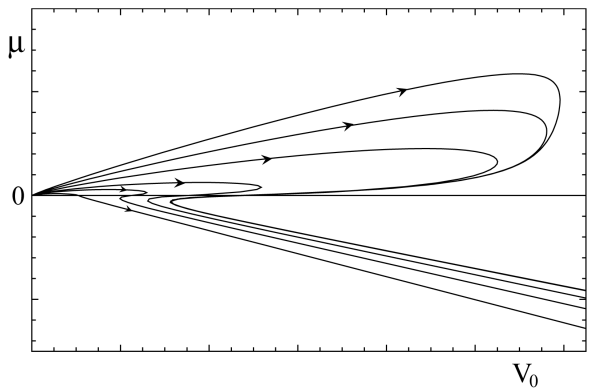}
\caption{\label{Fig.5} The RG flow diagram of the two sets of flow equations in the respective regions.}
\end{figure}

Second, there exists an attractor in the fourth quadrant towards which all the trajectories flow, without the need to fine tune the initial condition. The attractor can be found from solving the large-$t$ limit of the flow equations,
\beq\label{sf1}
\f{d\mu}{dt}&=&2\mu-\f{ \L^2 V_0 k_B T}{2 \pi} \f{2 \e_\L^3+10\e_\L^2\mu+4\e_\L\mu^2+2\mu^3}{[\e_\L(\e_\L+2\mu)]^2}=0,\nn\\
\f{d V_0}{dt}&=&2V_0-\f{ \L^2 V_0^2 k_B T}{2 \pi}\f{5 \e_\L^2+2\e_\L\mu+2\mu^2}{[\e_\L(\e_\L+2\mu)]^2}=0,
\eeq
which give $(V_0^*,\mu^*)=(0,-\e_{\L}/2)$. Due to a non-analytic behavior of the differential equations near the attractor, an expansion of \rfs{sf1} around the attractor $(V_0^*+\delta V_0 ,\mu^*+\delta \mu )$ gives
\beq
\f{d\delta\mu}{dt}&\simeq&-\e_\L+2\delta\mu +\f{9\L^2  k_B T \e_\L}{32 \pi}\f{\delta V_0}{\d \mu^2}\nn\\
&&-\f{15 \L^2 k_B T}{16 \pi}\f{\delta V_0}{\delta \mu}
-\f{\L^2 k_B T}{8 \pi\e_\L}\d V_0,\nn\\
\f{d \delta V_0}{dt}&\simeq&2\d V_0-\f{9\L^2 k_B T}{16\pi}\f{\d V_0^2}{\d\mu^2},
\eeq
which thus does not permit a linearization. The approach to the attractor can nevertheless be found by substituting the ansatz $\d V_0 \propto 1/t^2$. In this manner we find that the direction of approach is given by
\beq
\delta \mu^2 (t)\simeq \f{9\L^2 k_B T}{32 \pi}\f{t}{1+t}\delta V_0(t),
\eeq
and $\d \mu \propto 1/t$ for large $t$. This is shown in the inset of Fig. \rf{Fig.4}.

\begin{figure}
\includegraphics[scale=.6, angle=0, origin=c]{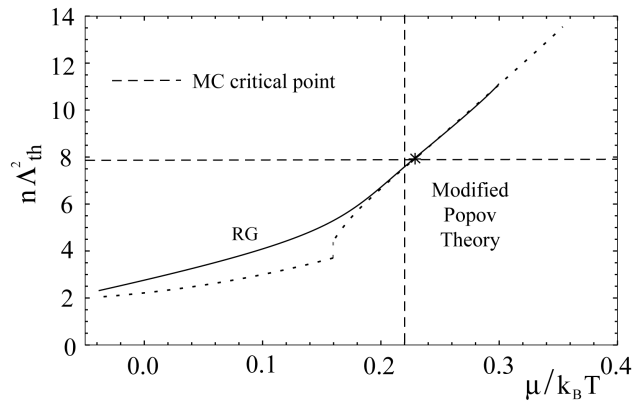}
\caption{\label{Fig.6} The RG equation of state, with $m g/\hb^2=0.15$. The dashed lines are the critical conditions for the BKT transition from the Monte Carlo (MC) results. The dotted line is the modified Popov theory.}
\end{figure}

Third, there exists a limiting curve that all trajectories approach asymptotically. The intercept of the trajectories with the $V_0$ axis for increasing chemical potential is shown in Fig. \rf{Fig.5a}. The interesting feature here is that for sufficiently large chemical potential the intercept essentially remains constant. We will come back to this point shortly.

Again, since the flow equations in the symmetry-broken phase are valid for $\mu>0$, the fourth quadrant should be considered as the unphysical region. However, the existence of an attractor and a limiting curve in the solution to the flow equations already suggests features which are unique to the ultracold 2D Bose gas, even though we did not take vortices into account explicitly. They are in fact related to the fixed line known from the BKT theory.

Finally, let us consider the system of coupled flow equations, and use the appropriate ones in the different quadrants. The result is shown in Fig. \rf{Fig.5}. The main point to note here is that a critical chemical potential $\mu_{c}$ can be identified for a fixed temperature, beyond which the resulting flow for large $t$ is insensitive to its initial value. This is so because for $\mu>\mu_c$, the trajectory changes the sign of $\mu(t)$ at essentially the same value of $V_0$, and thus the continuing flow into the fourth quadrant governed by the symmetric phase flow equations has essentially the same initial value. In other words, the long-wavelength action takes the same form for all $\mu>\mu_c$. As a result, we identify this $\mu_c$ with the critical condition for the BKT transition in this approach. It is interesting to note that the critical chemical potential obtained in this manner agrees within $10\%$ with the Monte Carlo result.

\begin{figure}
\includegraphics[scale=.71, angle=0, origin=c]{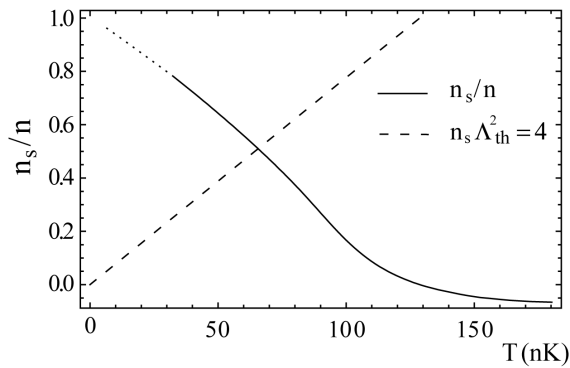}
\caption{\label{Fig.7} Superfluid fraction (solid curve) as a
function of temperature for a density of $1.5\times10^9
\textrm{cm}^{-2}$, with $m g/\hb^2=0.15$. The dotted line
extrapolates the solid curve from the RG calculation to the lower
temperature regime. The critical BKT condition is shown by the
dashed line.}
\end{figure}

\subsection{RG Equation of State and Correlation Effects }
Having discussed the properties of the flow equations, we next determine various nonuniversal quantities of interest. With the RG approach, the total density $n$ and the superfluid density $n_s$ can be computed by integrating these quantities along the trajectory, which are expressed by the following differential equations
\beq
\f{dn}{dt}&=&\f{\L^2}{2 \pi}\left[ \f{\e_{\L}+\mu}{2 \hb \o_{\L}}[2N(\hb \o_{\L})+1]-\f{1}{2}\right]e^{-2t},\textrm{ for } \mu>0, \nn\\
\f{dn}{dt}&=&\f{\L^2}{2 \pi}N(\e_{\L}-\mu)e^{-2t}, \textrm{ for }
\mu<0, \eeq and \beq\label{sfden}
\f{dn_s}{dt}=\f{dn}{dt}-\f{\L^2}{2 \pi}\e_{\L}\B N(x)\left[
N(x)+1\right]e^{-2 t}, \eeq where $x=\hb \o_{\L}$ or
$(\e_{\L}-\mu)$ for $\mu>0$ or $\mu<0$, respectively. The initial
conditions are $n(t=0)=n_s(t=0)=0$. In Fig. \rf{Fig.6}, the
density curve $n(t\rightarrow\8)$ as a function of the chemical
potential is shown and compared with the modified Popov theory.
For small chemical potentials, the small deviation from the
mean-field theory is consistent with the expectation that at low
densities the RG correction is unimportant. As the chemical
potential approaches the BKT critical point, however, the
deviation from the Hartree-Fock theory becomes substantial.
In fact, the RG density curve connects smoothly with the density
curve obtained from the modified Popov
theory. Thus, the RG approach resolves the artificial
discontinuity observed in the mean-field theory. Furthermore, it
shows that the equation of state for the quasi-condensate provides
already a good description above the critical temperature.

Next, by integrating \rfs{sfden}, the superfluid density is obtained, which is shown in Fig. \rf{Fig.7}. We see that the anticipated discontinuous jump in the superfluid density is absent at $n_s \L_{\textrm{th}}^2=4$, which shows that not all universal features of the BKT transition are incorporated yet. We believe that the effects on the superfluid density associated with the proliferating vortices can be taken into account explicitly by performing an additional renormalization group analysis on the Sine-Gordon model, as done in Ref. \cite{Stoof:02}. The initial condition for the dielectric constant $K(0)=\B\hb^2n_s/m$ should then be identified with the superfluid density obtained here.

\begin{figure}
\includegraphics[scale=.77, angle=0, origin=c]{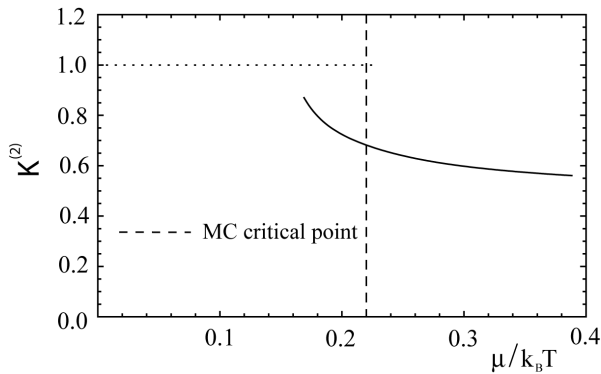}
\caption{\label{Fig.7a} Renormalized density-density correlator as a function of chemical potential, with $m g/\hb^2=0.15$. The dotted line is the prediction from the Hartree-Fock theory, with $\langle n^2\rangle=2\langle n \rangle^2$. The dashed line is the BKT transition from the Monte Carlo result.}
\end{figure}

\begin{figure}
\includegraphics[scale=.64, angle=0, origin=c]{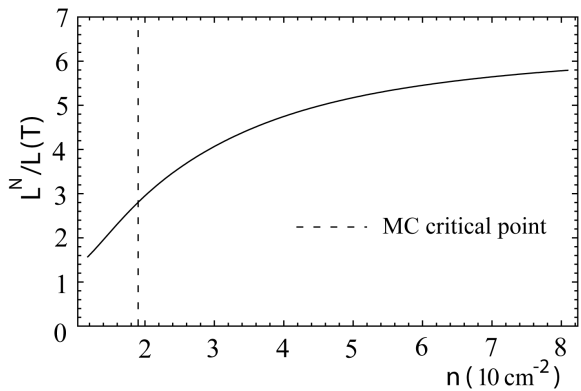}
\caption{\label{Fig.7b} Reduction of the three-body recombination rate as a function of the density at temperature $T=84.4 \textrm{ nK}$. The dashed line is the BKT transition from the Monte Carlo result.}
\end{figure}

Finally, we compute various many-body correlators where enhanced correlation effects are expected to show up as criticality is approach. From the modified Popov theory, the renormalized density-density correlator is given by \cite{Stoof:02}
\beq
K^{(2)}_R(T)&\equiv& \dl{\hat{\psi}^\dag(\tb{x})\hat{\psi}^\dag(\tb{x})\hat{\psi}(\tb{x})\hat{\psi}(\tb{x})}\rangle/2n^2\nn\\
&=&\f{1}{2 n^2}\left[ n_0^2+4 n_0 n'+2 (n')^2\right]. \eeq The
reduction in the three-body recombination is given by \beq
\f{L^N}{L(T)}\simeq\left\{ \left[\f{T_{00}(-2 \mu)}{T_{00}(-2 \hb
\Sigma)}\right]^6 K_R^{(3)}(T)  \right\}^{-1}, \eeq where $L^N$ is
the recombination rate constant in the normal phase, and the
renormalized three-body correlator is given by \beq
K_R^{(3)}(T)=\f{1}{6 n^3}\left[ n_0^3+9n_0^2 n' +18n_0 (n')^2+6
(n')^3\right]. \eeq In Fig. \ref{Fig.7a} and Fig. \ref{Fig.7b}, we
see that as the system in the normal phase approaches criticality,
the mean-field description of Hartree-Fock theory for the
many-body correlators breaks down completely. While the latter
gives a constant value of 1 for the renormalized density-density
correlator, the enhanced correlation due to the presence of a
quasicondensate reduces this value to 0.68 at criticality.
Furthermore, the three-body recombination rate reduces by a factor
of 2.8, compared to the recombination rate constant in the normal
phase. Thus, both these can serve as an observable for beyond
Hartree-Fock effects.



\section{Trapped Bose Gases}\label{sec:dis}
In this section, we extend our results to the inhomogeneous case of trapped Bose gases. For the case of a trapped \textit{ideal} Bose gas, the phenomenon of Bose-Einstein condensation (BEC) occurs. Within the local-density approximation (LDA), BEC takes place when the phase-space density in the center of the trap diverges. The temperature at which this phenomenon occurs, the BEC temperature $T_{BEC}$, can be related to the total particle number $N$ in the trap by \cite{Kleppner:91}
\beq
N=\f{\pi^2}{6}\left(\f{k_B T_{BEC}}{\hb \bar{\o}}\right)^2,
\eeq
where $\bar{\o}$ is the geometric mean of the radial trapping frequencies. For the case of a trapped interacting Bose gas in two dimensions, the system does not undergo a BEC, but a BKT transition if the trapping frequency $\bar{\o}$ is sufficiently low \cite{Bhaduri:00,Baym:07}. For current experiments of interest, the harmonic length $\sqrt{\hb/m\bar{\o}}$ associated with the radial trapping frequencies easily exceeds the de Broglie wavelength $\L_{\textrm{th}}$. Thus, the use of LDA by incorporating the effect of radial trapping through the introduction of a local chemical potential $\mu(r)=\mu-m \bar{\o}^2 r^2/2$ is readily justified. We show in Fig. \rf{Fig.8} the density profile of an ideal Bose gas at the critical condition and compare it with the case of an interacting Bose gas within the LDA. Here, we see the drastic effect of interactions in the 2D Bose gas. For these conditions, the RG density profile for the interacting gas is only slightly different from the Hartree-Fock theory, which is stable in this case. We include the density profile of an ideal classical Boltzmann gas for comparison.

\begin{figure}
\includegraphics[scale=.9, angle=0, origin=c]{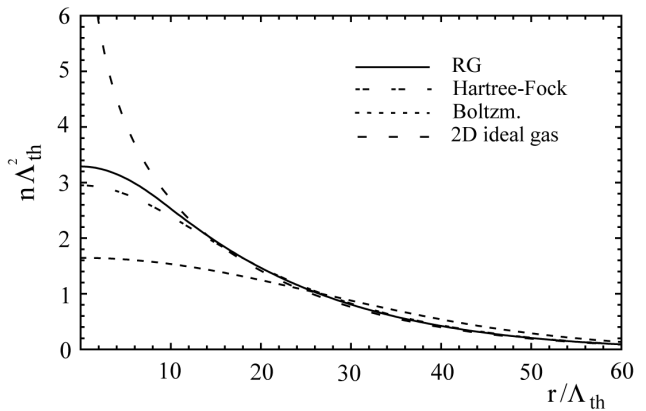}
\caption{\label{Fig.8} Density profiles for a quantum degenerate cloud of $^{87}\textrm{Rb}$ atoms for various theories for $N=7410$ at $T=T_{BEC}$. We take $a=5.2 \textrm{nm}$, a radial trapping frequency of $\bar{\o}=\sqrt{\o_x \o_y}=2 \pi \sqrt{9.4 \times 125}\textrm{ Hz}\simeq 2\pi \times 34.3 \textrm{ Hz}$ and an axial trapping frequency $\o_z=2\pi\times 4000 \textrm{ Hz}$. With this particle number, the phase-space density in the center of the trap diverges for the ideal Bose gas.}
\end{figure}
\begin{figure}
\includegraphics[scale=.87, angle=0, origin=c]{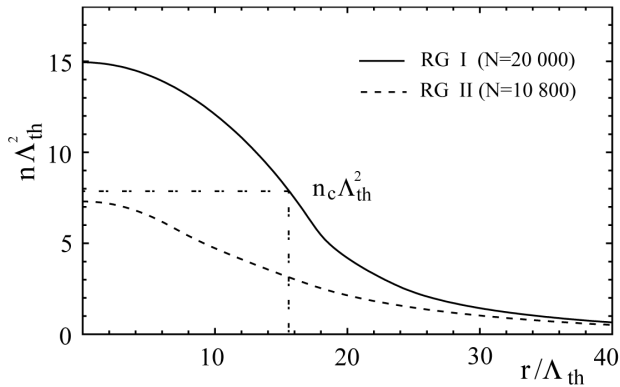}
\caption{\label{Fig.9} Density profiles from RG at $T=0.83 \,T_{BEC}$ and $T=0.61\, T_{BEC}$, for $N=10800$ and $N=20000$, respectively. In this range of phase-space density, the Hartree-Fock theory is unstable towards the equation of state of the modified Popov theory. The dash-dotted line is the critical condition for the BKT transition.}
\end{figure}

As the phase-space density is increased, the Hartree-Fock theory is no longer applicable due to the instability towards a more correlated state with a quasicondensate, as discussed previously. We show the RG density profiles in this range of phase-space density in Fig. \rf{Fig.9}. The two density profiles describe two different regimes of the gas, namely, one with the phase-space density in the center of the trap slightly below the critical phase-space density $n_c \L_{\textrm{th}}^2$ for the BKT transition, and the other well above it. We note that in the latter case, the gas consists of a superfluid core up to a critical radius, and is surrounded by an outer normal shell. One quantity of interest is the BKT transition temperature in the trap, relative to the ideal gas BEC temperature. By integrating the density profile at criticality, we obtain $T_{BKT}\simeq0.81 \,T_{BEC}$ for $mg/\hb^2=0.15$.

There has been recent interesting experimental and numerical work \cite{ Dalibard:07,Dalibard:07b,Krauth:07,Krauth:08,Blakie:08} being carried out to address the nonuniversal quantities of the quasi-2D trapped Bose gases. A direct comparison with our theory is rendered difficult, because in these works, the thermal excitations in the tightly confining direction are non-negligible.  Throughout our paper, we consider, instead, the strictly 2D regime, i.e., $\mu, k_B T \ll\hb \o_z$. \\

\section{Conclusion}\label{sec:con}
To conclude, we have studied various aspects of the 2D ultracold Bose gas.
First, we derived the exact form of the $T$-matrix for the 2D system as
realized in experiments. The 2D effective interaction assumes a form which
interpolates between the 2D and 3D results. We then presented the
mean-field results of the modified Popov theory. Even though the theory
can describe both the normal and superfluid states, the density turns out
to be discontinuous close to the BKT transition, which is an artifact of
the theory. We improved upon the mean-field description by a RG approach.
The flow equations exhibit interesting features, which resemble many of the
unique properties of the 2D $XY$ model, even though the effects of vortices
have not been included explicitly. With the RG approach, the density
correction to the normal equation of state indeed connects smoothly
with the quasicondensate equation of state in the superfluid phase.
We then computed various many-body correlators in the normal phase
close to criticality. We showed that deviations from the Hartree-Fock
theory are important, and they show up in the renormalized density-density
correlator and the reduction of the three-body recombination rate.
We finally extended the results to the inhomogeneous case of trapped Bose gases. The density profiles for various phase-space densities were evaluated and we found that close to criticality, the RG approach becomes necessary for a quantitative description of the gas. We hope that these beyond mean-field effects can also be observed experimentally in the near future.\\

\begin{acknowledgments}
We would like to thank R. Duine, K. Gubbels, and D. Makogon for
helpful discussions. We also thank J. Dalibard for suggesting us
to look more carefully at the density-density correlations.
\end{acknowledgments}

\appendix\section{}
\label{app}
In this appendix we describe in detail the derivation of \rfs{rTmat2}. Starting with \rfs{rTmat}, we add and subtract the quantity $\sum_n \int d^2 \tb{k} |\phi_n (0)|^2/(2 \pi)^2 E_0 (\tb{k},n) $ to obtain
\begin{widetext}
\beq\label{aeq1}
\frac{1}{T(E)}=\frac{m}{4 \pi \hbar^2 a}+ \f{m}{4 \pi \hb^2}\sum_{n=0}^{\infty} |\phi_{2n}(0)|^2\ln \left[ \f{4n+1}{4n-2E/\hb \o_z} \right]+\sum_{n=0}^{\infty} \int \frac{d^2 \tb{k}}{(2 \pi)^2}\f{|\phi_n(0)|^2}{E_0(\tb{k},n)}-\int \frac{d^3 \tb{q}}{(2 \pi)^3} \frac{1}{2 \epsilon_{\tb{q}}},
\eeq
\end{widetext}
where the zero-point energy $\hb \o_z/2$ has been subtracted from the energy argument $E$, and the integration
\begin{widetext}
\beq
\int \f{d^2 \tb{k}}{(2 \pi)^2} \left[ \f{1}{E+\hb\o_z/2-E_0(2n,\tb{k})}+\f{1}{E_0(2n,\tb{k})}\right]
=\f{m}{4\pi\hb^2}\ln\left[\f{4n+1}{4n-2E/\hb\o_z}\right]\nn
\eeq
\end{widetext}
has been carried out. The third term in the right-hand side of \rfs{aeq1} can be written as
\begin{widetext}
\beq\label{appen1}
\sum_{n=0}^{\infty} \int \frac{d^2 \tb{k}}{(2 \pi)^2}\f{|\phi_n(0)|^2}{E_0(\tb{k},n)}&=&\lim_{x\rightarrow 0}\sum_{n=0}^{\infty}\int  \frac{d^2 \tb{k}}{(2 \pi)^2} \f{H_{2n}(0)H_{2n}(x/\sqrt{2}l)e^{-x^2/4l}}{\sqrt{2\pi}l 2^{2n} (2n)! }\f{1}{E_0(\tb{k},2n)}\nonumber\\
&=&\lim_{x\rightarrow 0}\int  \frac{d^2 \tb{k}}{(2 \pi)^2}\f{e^{-x^2/4l}}{\sqrt{2\pi }l\hb \o_z}\sum_{n=0}^{\infty} \f{(-1)^n H_{2n}(x/\sqrt{2}l)}{2^{2n} n! }\f{1}{2n+\nu}\nonumber\\
&=&\lim_{x\rightarrow 0}\int  \frac{d^2 \tb{k}}{(2 \pi)^2}\f{e^{-x^2/4l}}{\sqrt{2\pi }l\hb \o_z}\sum_{n=0}^{\infty} \f{L^{(-1/2)}_n(x^2/2l^2)}{2n+\nu},
\eeq
\end{widetext}
where the variable $\nu=1/2+l^2\tb{k}^2$ is conveniently defined, $L^{(-1/2)}_n(x)$ is the generalized Laguerre function, and the relations $H_{2n}(0)=(-1)^n (2n)!/n!$ and $H_{2n}(x)=(-1)^n 2^{2n} n! L_n^{(-1/2)}(x^2)$ have been used.

Following Ref. \cite{Busch:98}, we use the integral representation
\beq
\f{1}{2 n+\nu}=\int_0^{\8}dy \f{1}{(1+y)^2}\left(\f{y}{1+y}\right)^{2n+\nu-1}\nn
\eeq
for $2 n+\nu>0$ and one of the generating functions of the generalized Laguerre function
\beq
\sum_{n=0}^{\infty}L_n^{(-1/2)}(x)z^n=(1-z)^{-1/2} \exp \left(\f{x z}{z-1}\right).\nn
\eeq
In this manner we find
\beq
\sum_{n=0}^{\infty} \f{L^{(-1/2)}_n(0)}{2n+\nu}&=&\int_0^{\8}dy\f{y^{\nu-1}(1+y)^{\nu-2}}{\sqrt{1+2 y}}
\nn\\&=&\f{\pi^2 2^{\nu-2} }{\G(1-\nu/2)^2 \G(\nu)  \sin^2(\pi\nu/2)}.\nn
\eeq
The last two terms in the right-hand side of \rfs{aeq1} can then be numerically integrated to give
\beq
&& \negthickspace\negthickspace\negthickspace\negthickspace\negthickspace\sum_{n=0}^{\infty}\int \frac{d^2 \tb{k}}{(2 \pi)^2}\f{|\phi_n(0)|^2}{E_0(\tb{k},n)}-\int \frac{d^3 \tb{q}}{(2 \pi)^3} \frac{1}{2 \epsilon_{\tb{q}}}\nn\\
&=&\f{\pi^{3/2}}{4\sqrt{2 }l\hb \o_z}\int \frac{d^2 \tb{k}}{(2 \pi)^2} \f{2^{\nu} }{\Gamma\left(1-\nu/2\right)^2 \Gamma(\nu) \sin^2(\pi\nu/2)}
\nn\\
&&-\f{m}{2\hb^2}\int\f{d^2\tb{k}}{(2\pi)^2}\f{1}{|\tb{k}|}\simeq -0.0279\,\,m/l \hb^2.
\eeq
Finally, by using $T_{00}(E)=|\phi_0(0)|^2T(E)$, \rfs{rTmat2} is obtained.


\begin{thebibliography}{100}
\bibitem{note1}
For a detailed derivation of the flow equations, we refer the reader to Ref. \cite{Stoof:96}.
\end{thebibliography}
\end{document}